\documentclass{article}

\usepackage{arxiv}

\usepackage[utf8]{inputenc} 
\usepackage[T1]{fontenc}    
\usepackage{hyperref}       
\usepackage{url}            
\usepackage{booktabs}       
\usepackage{amsfonts}       
\usepackage{nicefrac}       
\usepackage{microtype}      
\usepackage{lipsum}		
\usepackage{graphicx}
\usepackage{doi}
\usepackage{float}

\title{\emph{MAIScope}: A low-cost portable microscope with built in vision AI to automate microscopic diagnosis of diseases in remote rural settings}

\author{{Rohan Sangameswaran} \\
	Foothill College\\
	\texttt{20416638@fhda.edu} \\
}

\date{}

\renewcommand{\shorttitle}{\textit{MAIScope}}

\hypersetup{
pdftitle={MAIScope: A LOW-COST PORTABLE MICROSCOPE WITH BUILT IN VISION AI TO AUTOMATE MICROSCOPIC DIAGNOSIS OF DISEASES IN REMOTE RURAL SETTINGS},
pdfsubject={q-bio.NC, q-bio.QM},
pdfauthor={Rohan Sangameswaran},
pdfkeywords={First keyword, Second keyword, More},
}

\begin{document}
\maketitle

\begin{abstract}
According to the World Health Organization(WHO), malaria is estimated to have killed 627,000 people and infected over 241 million people in 2020 alone, a 12\% increase from 2019. Microscopic diagnosis of blood cells is the standard testing procedure to diagnose malaria. However, this style of diagnosis is expensive, time-consuming, and greatly subjective to human error, especially in developing nations that lack well-trained personnel to perform high-quality microscopy examinations. This paper proposes {Mass-AI-Scope (MAIScope)}: a novel, low-cost, portable device that can take microscopic images and automatically detect malaria parasites with embedded AI. The device has two subsystems. The first subsystem is an on-device multi-layered deep learning network, that detects red blood cells {(RBCs)} from microscopic images, followed by a malaria {parasite} classifier that {recognizes malaria parasites}
in the individual {RBCs.}
The testing and validation demonstrated a high average accuracy of 89.9\% {for classification and average precision of 61.5\% for detection models using TensorFlow Lite while addressing limited storage and computational capacity}. This system also has cloud synchronization, which sends images to the cloud when connected to the Internet for analysis and model improvement purposes. The second subsystem is the hardware which consists of components like Raspberry Pi, a camera, a touch screen display, and an innovative low-cost bead microscope. Evaluation of the bead microscope demonstrated similar image quality with that of expensive light microscopes. The device is designed to be portable and work in remote environments without {the I}nternet or power. The solution is extensible to other diseases requiring microscopy and can help standardize automation of disease diagnosis in rural parts of developing nations.	
	
\end{abstract}


\section{Introduction}
Malaria, a disease transmitted by mosquitoes, is estimated to have killed 627,000 people and infected over 241 million people in 2020 alone, a 12\% increase from 2019~\cite{bib1}. The disease is both preventable and curable yet is still the leading cause of death in many developing nations in Africa and Southeast Asia. Malaria is a huge problem in under-resourced countries with about 80\% of cases found in African countries and about 13\% in the Southeast Asia Region~\cite{bib2}. A large proportion of malaria-related deaths are in rural and tribal areas where access and utilization of health services is poor~\cite{bib2}. Eradicating malaria is a top priority for the World Health Organization {(WHO)} as well as governments of many under-resourced nations in Africa and Southeast Asia. However, practical methods of diagnosis and treatment are missing in rural areas which is one of the main reasons why this preventable disease has not been eradicated yet {~\cite{bib16}}.

According to the {Centers for Disease Control and Prevention (CDC)\footnote{https://www.cdc.gov/}}, the golden standard to diagnose malaria is to get a prepared slide of a patient’s blood, place it under a light microscope, and have an experienced laboratorian inspect the slide in search of the malaria parasite which is found in red blood cells~\cite{bib3}. However, this style of diagnosis is time-consuming and greatly subjective to human error especially in developing nations that lack well-trained personnel to perform high-quality microscopy examinations. An evaluation conducted on 184 laboratory personnel in Uganda to test the quality of field microscopy concluded with a trifling accuracy of 41\%\cite{bib4}, highlighting the lack of properly trained malaria microscopy experts in developing nations. The cost of training such personnel is also high\cite{bib3}, which is one of the reasons developing nations still have a large problem with the malaria parasite\cite{bib5}. Furthermore, the microscopes used to inspect malaria are often expensive, bulky, and require uninterrupted electricity, which is not guaranteed in rural areas.
{Malaria parasites can be quantified against Red Blood Cells (RBCs) or White Blood Cells (WBCs). To quantify malaria parasites against RBCs, we need to count the parasitized RBCs among total RBCs on the thin smear and report the results as \textit{Parasitemia} percentage.}
{Rapid Diagnostic Tests (RDTs) {are sometimes} used by clinicians to detect the {presence} of Plasmodium parasites in the human bloodstream; however, RDTs often cannot detect the parasites themselves at low concentration and quantify \textit{Parasitemia}. As a result, they often fail to assess infection severity and guide treatments in clinics~\cite{bib18, bib24}.}
{Therefore,} {both the}  existing microscopic and antigen-detecting rapid tests solutions are not completely suitable or practical for point of care testing (testing/diagnosis without the need for a laboratory), {especially for lower-density parasitemia that may occur in individuals without any symptoms while causing transmission ~\cite{bib23} }. 

{Furthermore, according to the WHO, the} Covid-19 pandemic has added more stress to already weak healthcare systems resulting in a large increase in malaria deaths from 2019 to 2020. Timely diagnosis of malaria {parasite} is critical as delays not only make the disease more dangerous for those infected but also increases the spread among the community and can often lead to widespread death. {So, effective, practical, and confirmatory diagnosis using microscopy technologies for global malaria control are urgently needed.}
Automating malaria diagnosis has significant advantages and has been a key focus of research{~\cite{bib17}}. Automation helps standardize interpretation of results, reduce diagnostic costs and helps malaria field workers be more resource efficient thereby serving more patients. There have been several efforts to detect malaria cells from microscopic images using deep learning and computer vision technology\cite{bib6,bib7,bib20}
{as well as}
developing low-cost microscopes\cite{bib8,bib9,bib18}. While these methods show great potential, they are often hard to test on a large scale, especially {at} point-of-care {treatment in} rural low resource settings, due to the difficulty {in replicating} their sophisticated hardware design or the need for expensive hardware to run the machine learning algorithm. For example, microscopes themselves cost hundreds of dollars. Moreover, there have been attempts to detect malaria with cloud infrastructure to bring automation to this problem, but this is not practical in rural areas as this type of solution requires a stable internet connection which can be expensive. Lastly, smartphone-based solutions {offer} {portable low-cost solutions{~\cite{bib21}}. However,  various software and hardware specifications and ever-changing camera options do not produce consistent results and keeping up with compatibility is impractical. }

This paper proposes a low-cost, portable device called MAIScope (Mass-AI-Scope) with embedded AI-based parasite detection technology to automate microscopy for malaria diagnostics. MAIScope is designed {to tackle} 
the practical constraints of {malaria diagnosis}
in rural areas. MAIScope’s innovation spans both the hardware and software domains. On the software side, it {is embedded with a multi-stage deep learning recognition pipeline} 
that directly runs on the device’s Raspberry Pi component without needing internet connectivity. 
{The pipeline} identifies and extracts individual {RBCs}
from the device’s microscopic imagery using its state-of-the-art blood cell object detection model. This is followed by running each 
{RBC} image through a deep image malaria {parasite} classifier trained on malaria microscopy data to detect malaria parasites. On the hardware side, the key innovation is to leverage low-cost bead microscope technology instead of expensive and bulky light microscopes. The hardware also uses other low-cost off-the-shelf components like a Raspberry Pi 4, camera, and touch screen display. The combination of innovative hardware and software design makes MAIScope an all-in-one microscopy automation tool which avoids the variability and inconvenience involved with piecing together independent microscopes with computer vision technology.\\ 

The Following are the key design criteria for the proof of concept being developed:
\begin{enumerate}

  \item Previously proposed solutions for the detection of malaria boast accuracy of 90\%+. Additionally, our proposed solution will run on a Raspberry Pi without internet access. Given these constraints, we set the minimum \textbf{expected accuracy} at 80\%.
\item The device needs to be \textbf{portable} for point-of-care testing (i.e. testing without the involvement of an expensive laboratory) in rural areas. The device size needs to fit into a 20x10x8 cm box and should be able to operate on battery and function without the {I}nternet.
  \item {The device} should be \textbf{easy to use} and intuitive to operate. A non-malaria expert should be able to scan and process a malaria slide within five minutes.
  \item The total cost of the device should be 
  \textbf{affordable} to distribute among a large rural population. The solutions that exist today involve expensive bulky microscopes that cost an average of \$500 or more. They also need expensive trained human experts and smartphones. The goal is to reduce the total cost by an order of magnitude and a \$125 cost was decided to be a very practical and effective goal. 
\item The overall design needs to be \textbf{extensible} such that the solution can be easily used in the future for other diagnostics use cases beyond malaria {parasites connected to RBCs}.
\end{enumerate}

\section{Prototype Construction}

{To provide a simple and effective point-of-care testing,}
MAIScope proposes a hardware implementation{(see Fig. 1 \& Fig. 2)} that consists of a small form factor {which is portable and easy to use in rural areas.}
Furthermore, MAIScope is designed to limit the amount of interaction with the user and in turn, limit the amount of human error involved with malaria diagnosis. To achieve this, the hardware user interaction (not including software) is limited to the following steps:  turn on the device, place a slide into the slide stage, and repeat until a screening session is done. To reduce cost, one of the core aspects of MAIScope’s hardware implementation is its low-cost embedded bead microscope \cite{bib10} providing a cheap yet effective microscope while avoiding the size and immobility of conventional Light microscopes. This is done with the use of a tiny glass bead
which{,} when implemented with a camera properly, can provide the camera image with magnification between 100x all the way to 1000x depending on the size of the bead (more {detail is provided in 
Embedded Bead Microscope section}). Pairing this with other low-cost hardware provides an all-inclusive, portable malaria screener. {Our design mainly consists of the following components to address the performance, portability and cost requirements.}

\begin{figure}[H]
\centering
\includegraphics[width=0.6\textwidth]{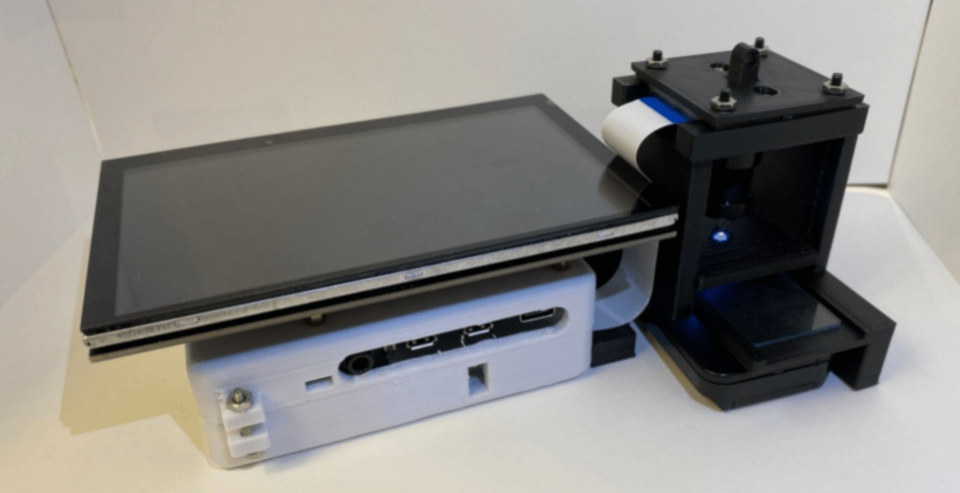}
\caption{MAIScope Hardware prototype with Case.}
\label{fig:hardware_with_case}
\end{figure}

\begin{figure}[H]
\centering
\includegraphics[width=0.6\textwidth]{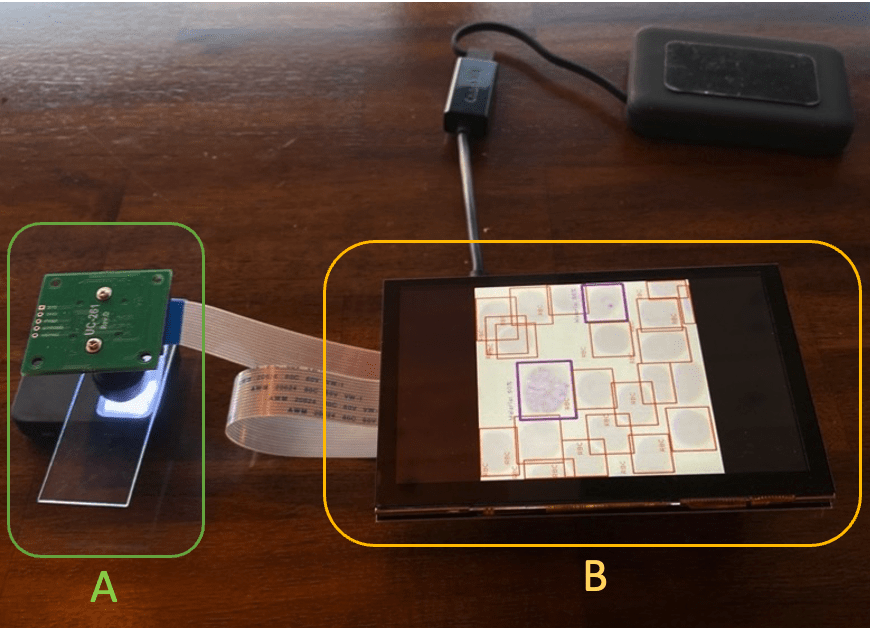}
\caption{{MAIScope prototype with no Case. Part A shows Raspberry Pi camera API (pi Camera) and part B shows Raspberry Pi LCD Display (Tkinter GUI).} }
\label{fig:hardware_2}
\end{figure}

\subsection*{Raspberry Pi}
The computer chosen for the MAIScope is the Raspberry Pi 4 Model B (Pi), it has a quad-core, Broadcom BCM2711 processor and can have between 2GB-8GB of RAM. The Pi has a footprint of only {$5.4\;cm^{3}$}
and offers similar performance to many entry-level laptops.
{Moreover, }the Pi has ports that enable it to connect to displays and cameras which gives MAIScope the ability to capture images and display them to the user. The Pi also has {WiFi} 
connectivity which enables MAIScope to send data to the cloud if internet is available. The price of the Raspberry Pi 4 is also cheaper starting at \$35 compared to other small form factor computers such as Nvidia's Jetson Nano which starts at \$60. For MAIScope, the Pi is used to {run}
the blood cell object detection 
and 
malaria {parasite} classification {models}. To power the Raspberry Pi, a portable battery pack 
or any other source of power that outputs 5V of power is necessary{.} 
The size of the battery {pack}
depends on how long the screening session will last.

\subsubsection*{Embedded Bead Microscope}
For MAIScope to detect malaria, the camera must see an image that is magnified at about 400x. 
To achieve this at a low-cost, MAIScope uses a glass bead with a diameter of 1mm. This style of the microscope is known as Leeuwenhoek’s microscope \cite{bib11}, and it works due to the spherical shape of the bead bending light in a manner that results in a magnified image. The advantage of the bead microscope is its small form factor and cheap price compared to conventional light microscopes used for microscopy. Due to the bead microscopes’ small form factor, MAIScope is portable and enables {point-of-care}
diagnoses and treatment while still giving a powerful magnification of 350x at a low-cost. This style of microscope was repurposed by researchers at the Pacific Northwest National Laboratory for use with mobile phones\cite{bib10}. It has since not been widely adopted in research {settings}
{because the ``eyepiece'' in this microscope is often extremely tiny and is only the diameter of the glass bead which can be viewed properly with a camera and not with the human eye.
}

\subsection*{Camera}
To capture images through the glass bead microscope, MAIScope has an 8MP Raspberry {Pi}
Camera with a 16mm lens attached to it. The 8MP camera offers a high definition 1080p image and is paired with a 16mm lens which gives the camera almost a 10x zoom. The use of a lens avoids the issue of digital zoom which often leads to a pixelated images that can hinder the machine learning models. Along with this, the microscope is still contained within a small form factor. The high definition and zoom paired with the embedded bead microscope allows the user to take high-quality images of a blood slide at high magnification, enabling the model to perform accurately as if it was looking through a conventional light microscope. 

\subsection*{Touchscreen Display}
MAIScope uses a 5-inch LCD touchscreen display which gives the user a mobile phone-like interface and removes the need for a keyboard and mouse to control the raspberry {Pi.}
This LCD display is also not very large with dimensions of 12.1 cm by 7.6 cm maintaining the portability for the MAIScope. This size is comparable to a mobile phone screen.

\subsection*{Case, Slide Stage, and LED}
MAIScope has a case that houses the Raspberry Pi and provides it {with} a degree of protection. The case also has the Microscopy module{,} which houses the camera mount and the slide stage{,} attached to it. 
The camera mount keeps the camera in place such that it looks down at an inserted slide on the slide stage and is screwed so that there is no unnecessary camera movement. The slide stage consists of a piece of plastic with a groove in it in which a slide can be placed. Below the slide stage is space for an LED that passes light through the microscope slide like a conventional light microscope, illuminating the image for clear viewing. {A 3D design of MAIScope is shown in Fig. 3 and Fig. 4}

\begin{figure}[H]
\centering
\includegraphics[width=0.98\textwidth]{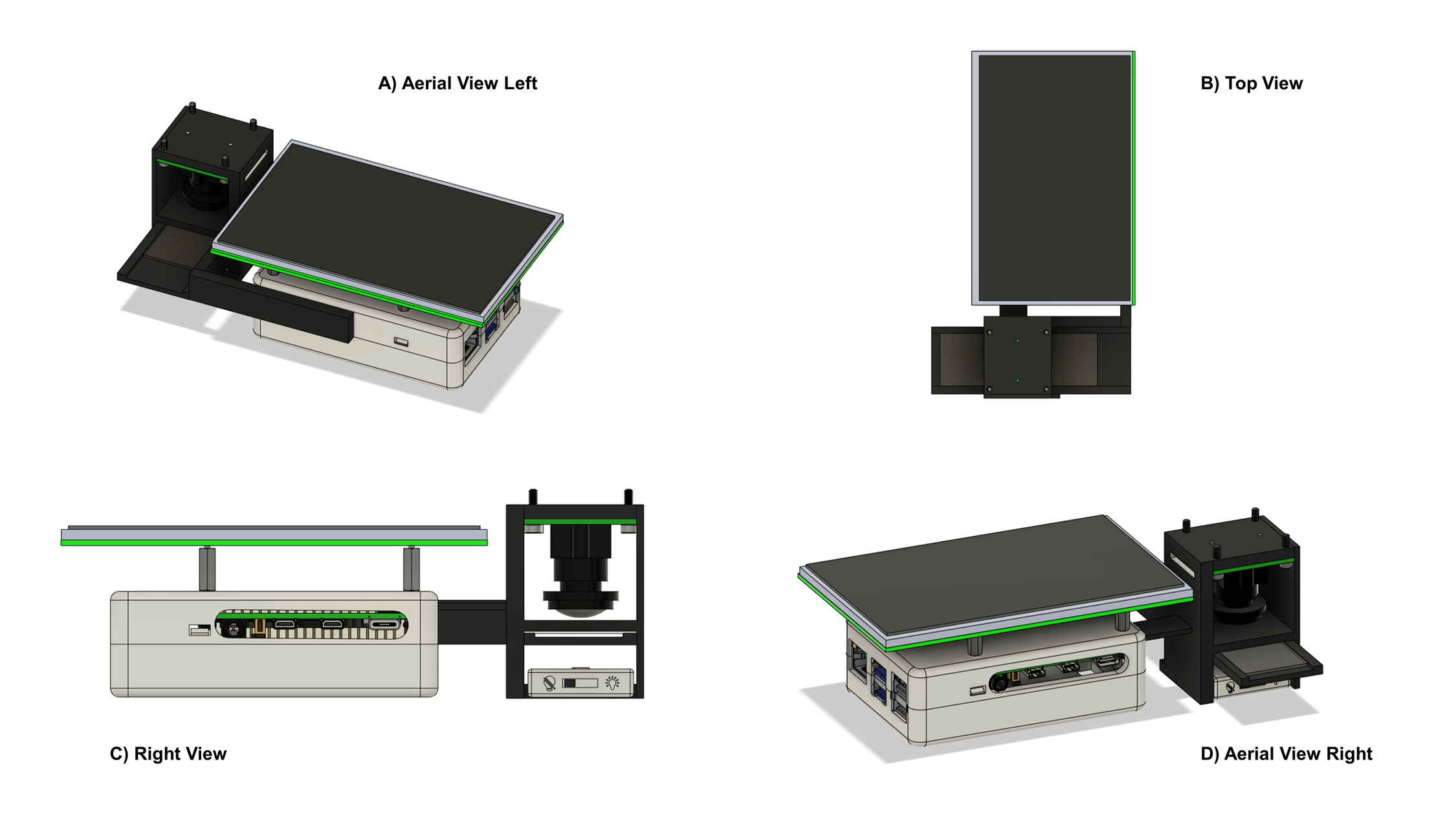}
\caption{3D design of MAIScope made in Auto Desk Fusion 360.}
\label{fig:hardware_1}
\end{figure}

\begin{figure}[!h]
\centering
\includegraphics[width=0.98\textwidth]{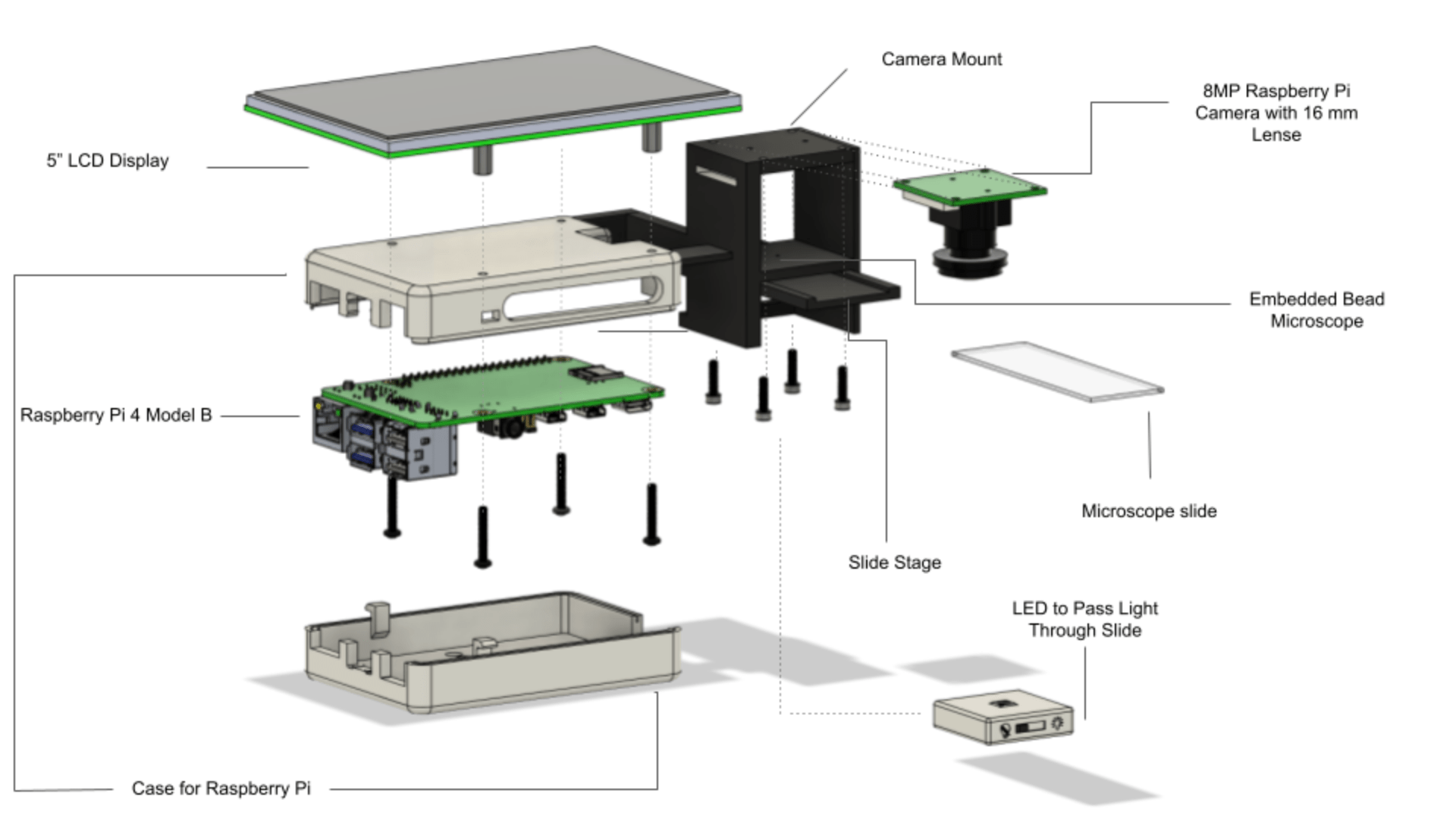}
\caption{Exploded view of 3D design of MAIScope made in Auto Desk Fusion 360.}
\label{fig:hardware_2}
\end{figure}

\section{ {Embedded Software}}
{The software workflow of MAIScope is shown in Fig 5. The image capturing module captures an image of the magnified slide from the Raspberry Pi camera feed. The image is then passed into the malaria identification pipeline including blood cell detection and malaria classification models, explained in the following sections in detail. The results are then displayed on the touch screen LCD for the user. The captured data and results are stored locally if the user later chooses to send them to the cloud platform for further analysis once MAIScope connects to the Internet.}

The machine learning models, and pipeline take into account the following constraints:
\begin{enumerate}
  \item On-device low latency run-time due to lack of internet connectivity
  \item Small model size to run on embedded device
  \item Low power consumption and efficient inferencing for longer battery life 
\end{enumerate}
The first approach was to develop an image classifier that classified the entire microscopic blood slide image for 
{malaria parasite recognition}. However, this uncovered several issues. For example, 
{WBCs} and platelets looked very similar to malaria{-infected} cells and malaria only infects 
{RBCs}. 
{Apart from leading to many false positives,} this approach also could not count the number of cells infected by malaria which is also important to understand the stage of spread and development {(\textit{Parasitemia})}.

{To overocme this, a multi-stage deep learning pipeline(shown in Fig. 5) was developed that first detects and identifies RBCs in the slide image. Each of the detected RBCs are then cropped to send to the malaria parasite classification model for parasite prediction. This multi-stage pipeline was done to eliminate WBCs and platelets which can interfere with the malaria parasite recognition algorithm in the classification stage. The pipeline also keeps count of the total number of found RBCs as well as the number of infected cells by malaria. Furthermore, in the event that a single cell from an image is misclassified, the entire image won't wrongly be classified as being parasite free.}

\begin{figure}[!h]
\centering
\includegraphics[width=1\textwidth]{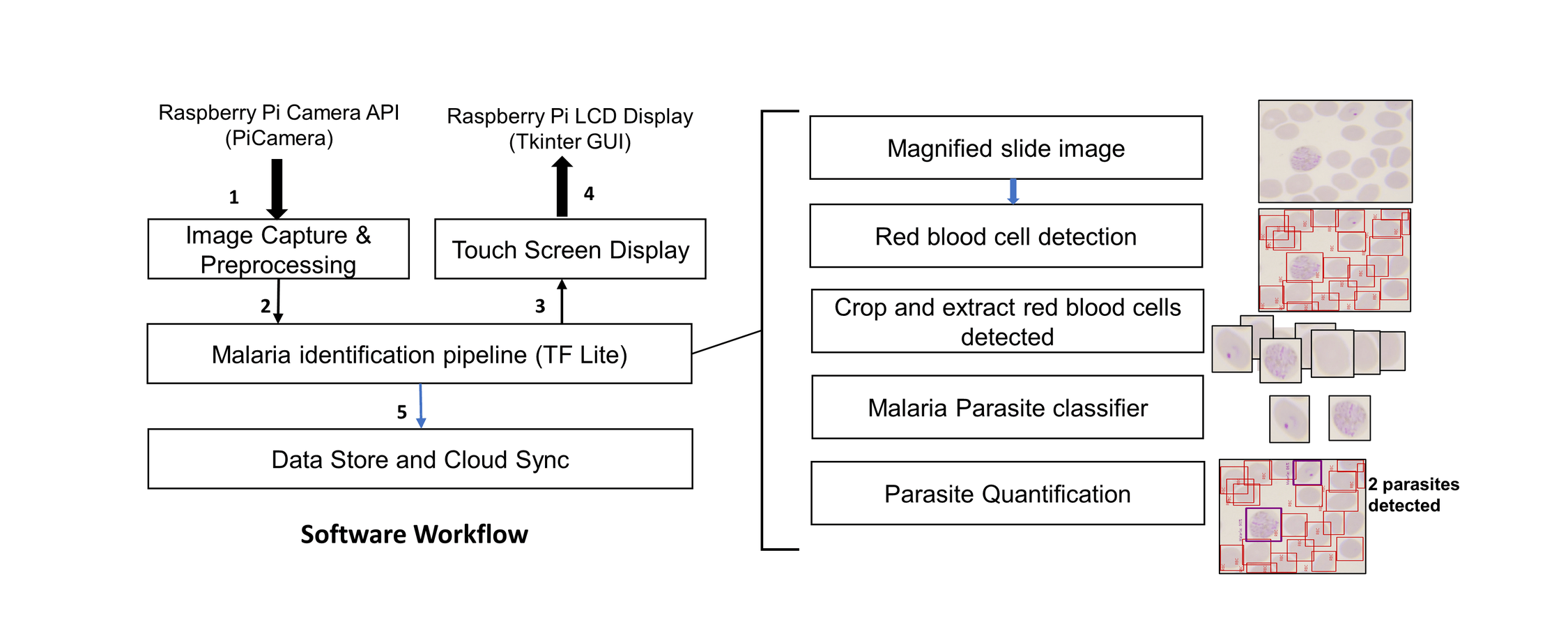}
\caption{The five steps in MAIscope Software workflow which are used to identify malaria from a sample.}
\label{fig:software_wf}
\end{figure}


\subsection*{{Image Capturing and Pre-Processing}}
{The first part of our embedded software is the image capturing module. It uses the Raspberry Pi’s camera interface to capture an image from the camera looking at a magnified slide. After the image is captured, it is resized to 320x320 using the OpenCV re-sizing module since the blood cell detection model requires an input image of these dimensions.}

\subsection*{{Blood Cell Detection}}
The goal of our blood cell detection model is to locate and label the cells as RBCS, WBCs or platelets. 
This requires a dataset with a large {number of}
images that are annotated {for object detection tasks (i.e., having labelled objects with rectangular or bounding boxes).}

Such datasets are not widely available {and are not easy to construct}. {Therefore,} a small-scale dataset {named} BCCD \cite{bib11} {(shown in Fig. 6)} was used to bootstrap the model. 
The dataset includes 364 images of thin smear microscopic slide images across 3 classes - RBCs, WBCs, and Platelets. Since the dataset is not very large, several data augmentation techniques (rotation, resizing,hue, etc.) were used during training.


\begin{figure}[!h]
\centering
\includegraphics[width=0.6\textwidth]{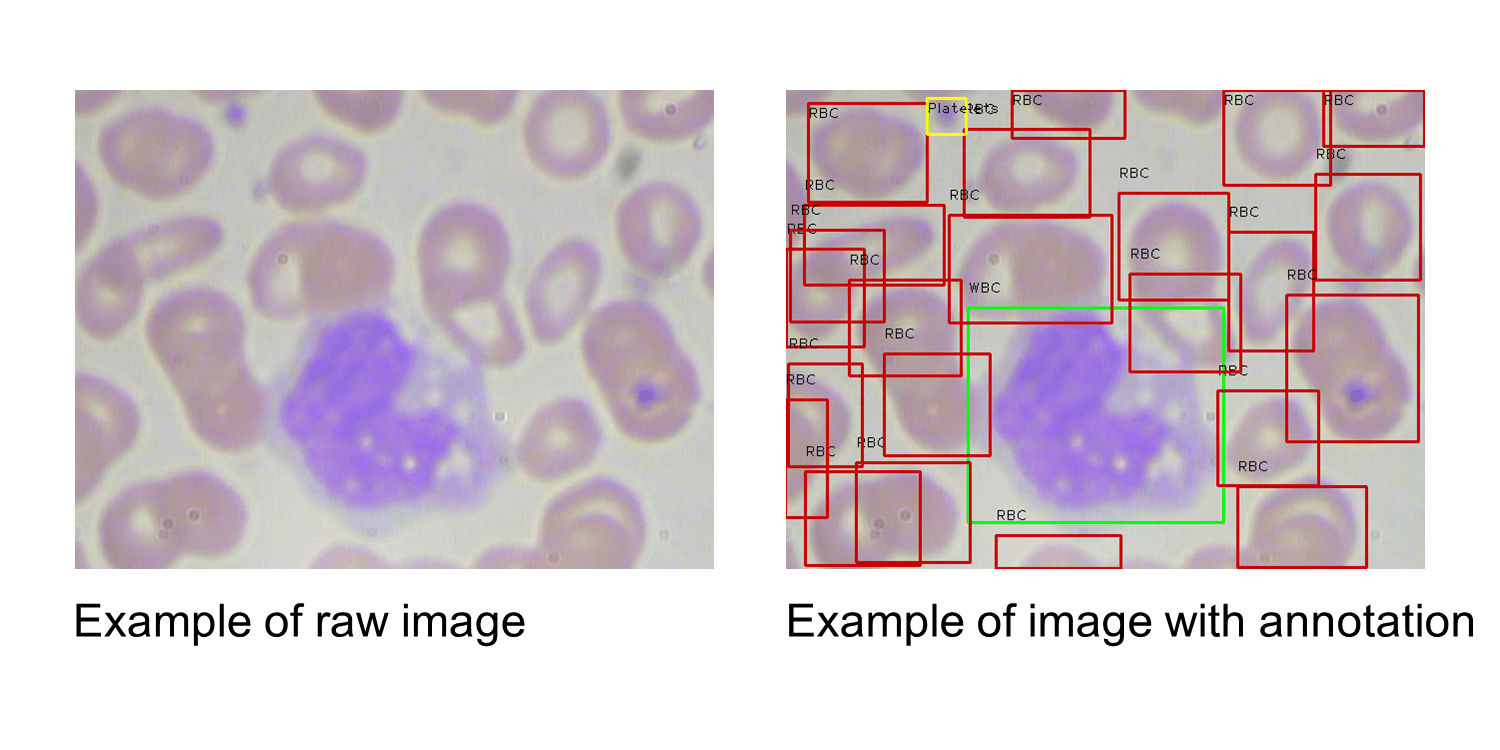}
\caption{{Sample data from BCCD Data used to train blood cell detection model. The green rectangles show the RBCs, the red ones show WBCs, and the purple ones show the Platelets. }}
\label{fig:software_wf}
\end{figure}


Since the 
{BCCD} dataset is a small, training 
an object detection model {from scratch} would not give the kind of accuracy needed to provide proper cell detection.
{Due to this reason,}
transfer learning{~\cite{bib26}} was used 
{as a machine learning} technique 
that focuses on storing knowledge gained while solving one problem and applying it to a different but related problem.

TensorFlow Lite, a version of TensorFlow designed for mobile and IoT devices, was used as the platform to develop the machine learning models. The deep learning object detection architecture chosen was EfficientDet-lite which uses layers of convolutional neural networks and scaling; there are 5 versions, as 
{compared in \cite{bib25}},
of this architecture: EfficientDet-lite[0-4] with Efficientdet-lite0 being the baseline architecture and the rest being scaled-up versions. The EfficientDet-lite models are {pre-}trained on the COCO dataset \cite{bib12} for object detection.

This base model was frozen such that all the inner layers were frozen and the last few classification layers were fine tuned for the BCCD dataset. This was done so that the features learned during object detection with the COCO dataset can then be used for blood cell object detection.

For MAIScope, of the different variations of Efficientdet-lite models, Efficeintdet-lite4 was used as it provides the highest accuracy compared to the rest.
This model also can still run efficiently on low-end hardware. Even though EfficeintDet-lite4 has the highest latency compared to the other models, MAIScope needs to achieve the highest accuracy while still being efficient enough to run on low-end hardware and Eficientdet-lite4 does just that. This object detection model has an input of a 320x320 pixel image and has 3 outputs: class, score, and location. The class is the identified class of an object and for the blood cell detection model the classes are 
{RBCs, WBCs}, and Platelets. The score is the {probability score}  of the model and is represented as a decimal between 0 and 1.
The location output is the location of the identified object and is given as an array of coordinates [top, left, bottom, right]. Using the values provided by the output of the model, the output can be visualized with an overlay in which bounding boxes with class labels and percentage confidence can be placed over objects in the original image giving the user a simple visualization of the output. The workflow 
{of our} blood cell detection model is shown in Fig. 7 below.

\begin{figure}[H]
\centering
\includegraphics[width=0.86\textwidth]{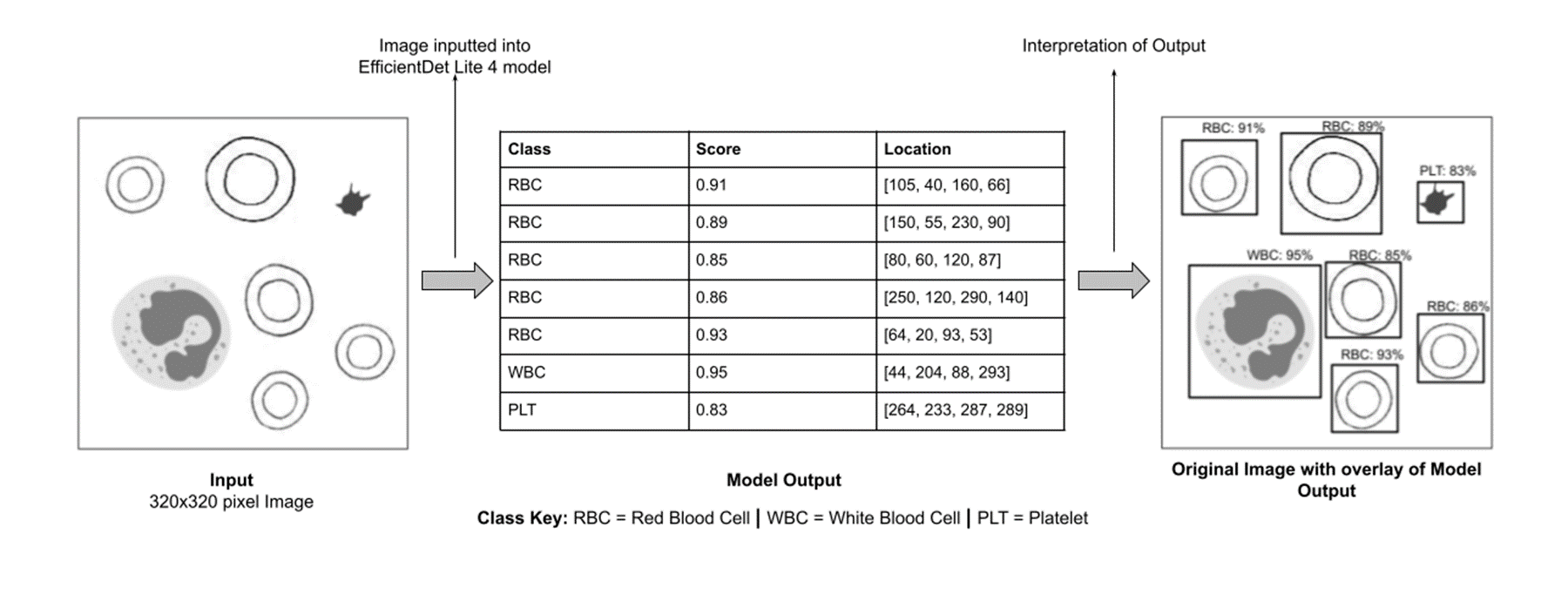}
\caption{Workflow for MAIScope blood cell detection model.}
\label{fig:software_wf}
\end{figure}


\subsection*{{Malaria Parasite Classification}}
In order to detect the malaria parasite in red blood cells with high accuracy, the model needs a large, well-formatted dataset of malaria-infected and uninfected red blood cells. In this case, a dataset from the National Library of Medicine \cite{bib13} was used which consists of 27,558 images of blood cells with half being infected and the other half being uninfected. Fig. 8 shows examples of this dataset. Additional data augmentation was {used} to increase the number of training images and expose the model to a variety of different-looking cells. 


\begin{figure}[!h]
\centering
\includegraphics[width=0.7\textwidth]{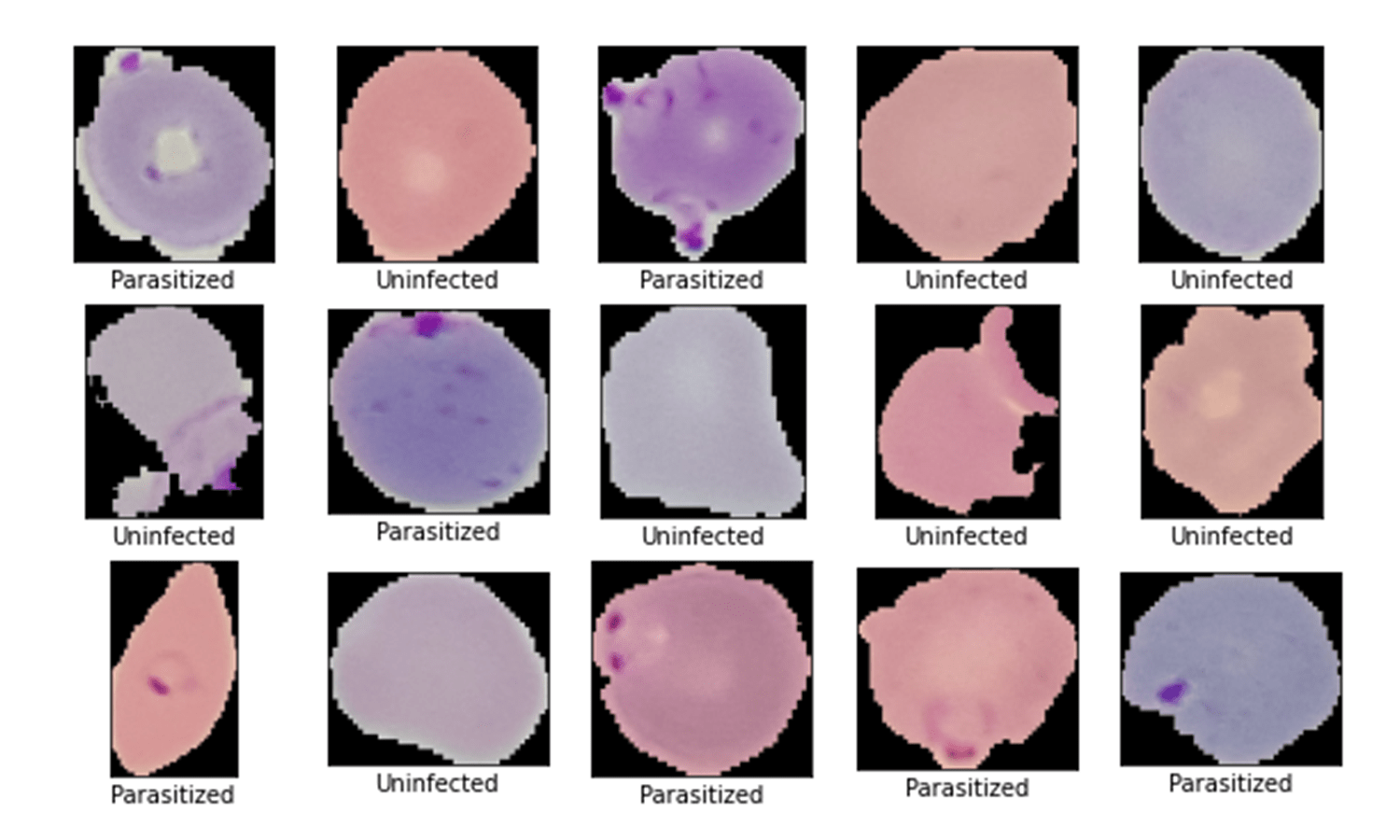}
\caption{{Sample images used to train the malaria cell classification model from the National Library of Medicine~\cite{bib13}.}}
\label{fig:software_wf}
\end{figure}

The malaria cell classification model also uses TensorFlow Lite 
{as well as } Transfer Learning techniques.
MobileNetv2, a high-performance Convolutional Neural Network model optimized for mobile devices was chosen as the classification model architecture. This model is {pre-}trained on a very large general image dataset called ImageNet\cite{bib14}. This convolutional base model was then frozen such that all the feature learning from this model can be applied to the final layer trained on the custom malaria dataset described above so that it can classify cells that are infected with high confidence. The model was then quantized, a machine learning technique that decreases model size with minimal impact on accuracy. This was done to make sure the model runs efficiently while running on the Raspberry Pi 4. The malaria cell classification model takes in a 224x224 pixel image and after passing the image through the model, outputs
{a probability distribution across infected and uninfected classes}. 
The workflow for the malaria classification model is shown in Fig. 9.

 \begin{figure}[H]
\centering
\includegraphics[width=0.75\textwidth]{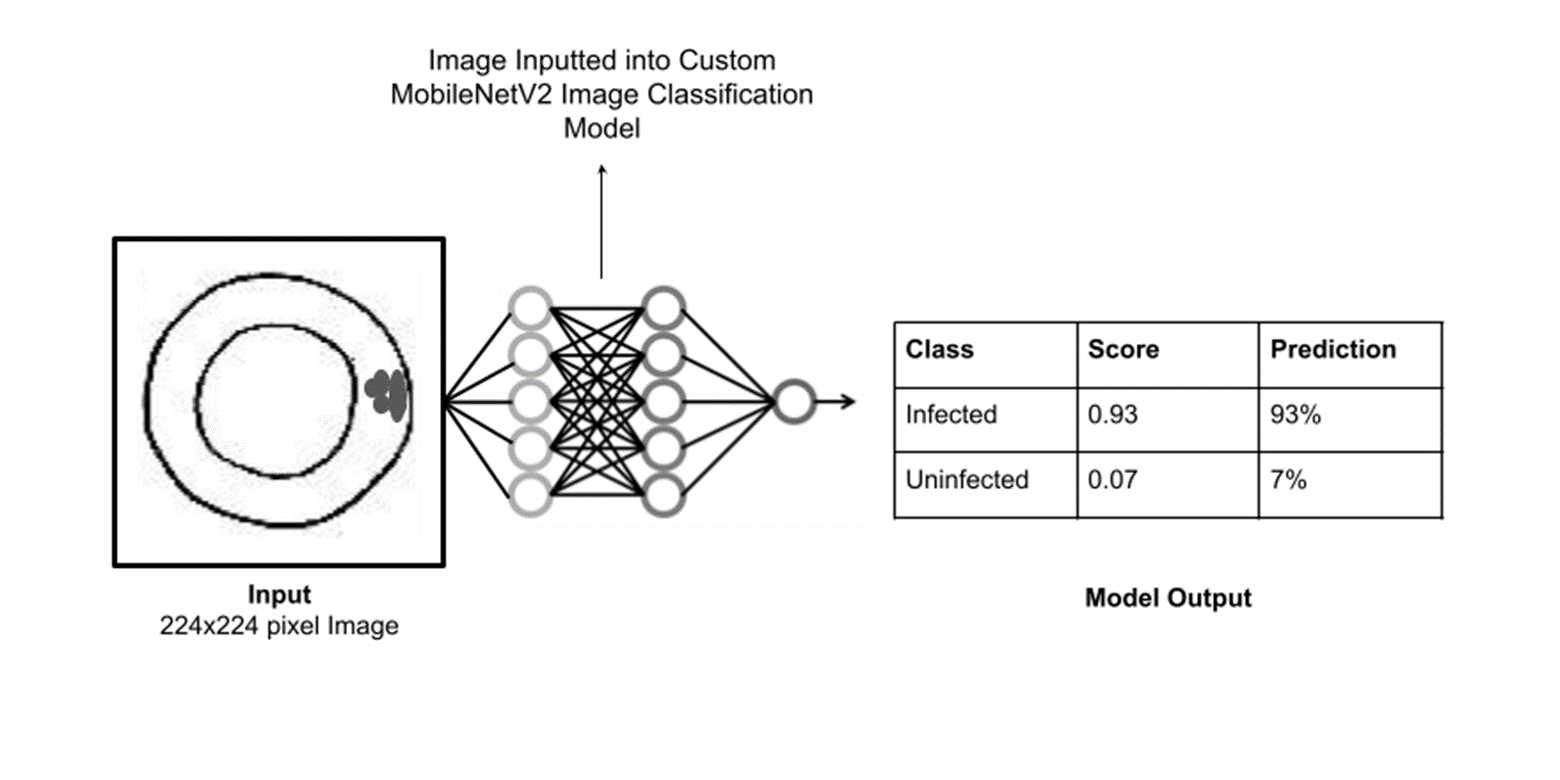}
 \caption{Malaria {parasite} classification model workflow.
 }
 \label{fig:software_wf}
 \end{figure}


\subsection*{{Malaria Parasite Recognition Pipeline}}
After creating 
{our} blood cell detection and malaria {parasite} classification 
models, they were combined into a pipeline where the output of 
{the former one} 
is used as the input 
{of the latter one.} In this case, the original blood slide image is passed through the blood cell detection model, and every identified 
{RBC} is cropped, resized to 224x224 pixels, and is passed into the malaria {parasite} classification model. The extracted cells that are passed into the 
model are saved to the device for later review if necessary. More on this in the Data Storage and Cloud synchronization section. If the 
classification model output prediction is greater than 80\% for the infected class, the bounding box around the cell will be purple and the label will say, 
{``Malaria''} instead of ``RBC''. Otherwise, the bounding box around each object will be that of which is 
{generated} by the 
object detection model and its corresponding class label. The reason why 
{RBCs} are the only objects extracted from the 
detection model is that the malaria parasite only infects 
{RBCs} and passing any other type of cell could throw off the model. The two models are combined because simply passing a whole slide image into a malaria classification model can easily be thrown off by the presence of 
{WBCs} and platelets due to their similarity in color to the malaria parasite. Furthermore, counting the number of infected and uninfected cells would not be possible because the output would simply be percentage confidence. {We apply two separate models (i.e., detection and classification ones) in the pipeline instead of an End-2-End (E2E) training fashion (i.e., training the detection part with 4 classes including infected RBC, uninfected RBC, WBC, and Platelets). In such manner, in addition to dealing with data scarcity for the E2E model training, the classification part gets benefits from training on a large-scale available dataset, from the National Library of Medicine, although there is a shortage of data for the detection part, as we only have the small-scale BCCD dataset.}



\subsubsection*{Parasite Quantification}
{
One of the outputs of the model that is provided to the user and saved on the device is a count of the number of 
infected and uninfected RBCs. This is an important metric of disease severity and can influence the choice of treatment options.}


\subsubsection*{Touch Screen Display Module}
The touch screen display module uses a Tkinter Graphical User Interface (GUI) 
to show the user a visualization of the model pipeline’s final results. In this GUI, the user can decide when to take photos to run through the malaria Identification pipeline. It is also used to control the MAIScope as there are no other peripherals for ease of use. 

\subsubsection*{Data Storage and Cloud Synchronization}
This module stores the slide and cell images and predictions in a local database. Once the device connects to the Internet, the slide image data is synchronized to the cloud where all the data is stored in a database. This data can then be used to further fine tune the model for {a} better accuracy. This is especially useful since there is not enough slide data with field stains in rural areas available. Such data is key to enhancing the machine learning model.

\subsection*{User Experience}
MAIScope’s user experience is intuitive, designed with ease of use in mind. Following are the steps that a user would follow to screen for malaria.

\begin{enumerate}
  \item Switch on the device and LED. This will boot up the device and launch the software user interface
  {to indicate that it is} 
  ready for the slide to be screened.
  \item Insert prepared blood slide into the slide stage on the device. The LCD display will show the live preview of the magnified blood slide coming through the embedded bead microscope and camera. 
  \item The user clicks the “capture image” button on the touch screen display.
  \item The image is then processed and the results are displayed on the screen which include the infected and uninfected cell count as well {as} an overlay of purple bounding boxes around malaria infected cells.
  \item The user can close the screen and will be given a choice to save the slide, individual cell images and results for later use. Steps 2 through 5 are repeated for the next slide screening.
\end{enumerate}

\section{Results}
{We run our TensorFlow Lite-based models on the Raspberry Pi, as a low-cost and high-performing computer. The following sections explain our evaluations results for both detection and classification tasks in such setting.}

\subsection*{Testing and Validation of the Blood Cell Detection Model}
The BCCD dataset, which consists of microscopic images of blood components along with annotations specifying the location and type of each cell in each image, was split into three parts with 70\% percent of the data being used for training and the remaining 30\% as validation and testing. In addition to the above, manual testing was done with images of field microscope slides with malaria from Cancyte{\footnote{https://www.cancyte.com/}} labs (see testing of model pipeline on real world data) to validate 
{how}
the model 
{performs}
with real world data.

{Table 1} compares MAIScope’s blood cell object detection model to other model architectures trained on the BCCD dataset. {Mean average precision (mAP) is calculated for assessing the object detection models (we use AP for the sake of simplicity). Based on the ground truth, a higher AP means that objects are detected more accurately. AP can be also calculated either with a specific IoU (i.e., Intersection over Union as the overlap between ground truth and detected bounding boxes) threshold (we report $AP_{50}$ and $AP_{75}$ with $50\%$ and $75\%$ thresholds respectively) or for a range of specific object sizes (we report $AP_{S}$, $AP_{M}$ and $AP_{L}$ describing AP over small, medium and large objects respectively). We also report $AP$ which itself is the average of 10 APs having IoU thresholds of $0.5:0.95$ with steps of 0.05. As a result, we report 6 evaluation metrics to compare the models and provide detailed analysis.}

YOLO{v3~\cite{bib27}}
 {and} YOLOF{~\cite{bib28}}
 models are 
 designed to be used on more expensive hardware, hence the number of parameters is so much higher compared to MAIScope’s blood cell detection model. 
{The results show that} MAIScope’s model outperforms {these}
models  {while they have much larger model sizes as well as YOLOF+ with a same size using 4 metrics out of 6 ones.}
{The model can generate better results with different threshold values (i.e., $AP$, $AP_{50}$ and $AP_{75}$) showing comprehensive improvements. It can also generate better results on large objects ($AP_{L}$), but not on small ($AP_{S}$) and medium ($AP_{M}$) ones. YOLOF and YOLOF+ are more successful on detecting these sizes while MAIScope’s model generates comparable performance against YOLOv3.}

\begin{table}[H]
	\caption{Evaluation of MAIScope's blood cell detection model}
	\centering
	\begin{tabular}{llllllll}
		\midrule
		\bf{Model}     & \bf \#params     & \bf $AP$ & \bf $AP_{50}$ & $AP_{75}$ & $AP_{S}$ & $AP_{M}$ & $AP_{L}$ \\
		\midrule
		YOLO v3 \cite{bib29}  & 62M & 45.9 & 86.7 & 45.3 & 22.6 & 50.4 & 38.8    \\
		YOLOF \cite{bib29}     & 42M & 57.5 & 89.0 & 65.7 & 43.8 & 57.9 & 47.3 \\
		YOLOF+ \cite{bib29}  & 42M & 53.1 & 87.6 & 56.2 & 36.3 & 56.3 & 45.4\\
		\textbf{MAIScope Efficientdet-lite4} & \textbf{15M} & \textbf{61.5} &  \textbf{89.4}& \textbf{68.6} & {25.9} & {48.5} & \textbf{50} \\
		\bottomrule
	\end{tabular}
	\label{tab:table}
\end{table}

\subsection*{Testing and Validation of the Malaria {Parasite} Classifier}
The malaria dataset, which consists of {RBC}
images labeled infected or uninfected is used to train the MobileNetV2 classifier as discussed in {the malaria parasite classification}
section. The data was split 80-10-10 between training, testing and validation. The results are shown below. Additional manual tests were done on the malaria slide images obtained from Cancyte labs to test the end-to-end usability and accuracy.

Table 2 below compares MAIScope’s MobileNetV2 
model to other models trianed on the same dataset designed for individual malaria {parasite} cell classification. The malaria {parasite} classification model used in MAIScope is able to achieve high accuracy of 89.9\% on test data on a per detected cell basis while still being efficient enough to run on {a low-end}
hardware. This accuracy meets the design criteria from the beginning of the project.
The custom ResNet-50 Xception and AlexNet model are other state of the art models\cite{bib32} designed to classify malaria cells and were chosen as comparisons because they are trained on the same dataset and are common architectures used for image classification. These state of the art models, perform slightly better in regard to test accurarcy with the largest disparity being 6\% between the ResNet-50 model and MAIScope's model. However, MAIScope is the smallest model compared to the rest with 2.2 million paramaters which is 10x smaller than the AlexNet model, which is the smallest of the models being compared.  This data highlights how MAIScope’s malaria classification model is comparable to models that require high scale computing hardware while achieving similar or better performance at a fraction of the price on low-end edge hardware device. Furthermore, the 
classification model file size is only 2.7 MB while other computer vision models are often in the tens or hundreds of megabyte file size. It's very critical for model sizes to be small to be able {to deploy and }run efficiently on the device without running out of RAM/Memory.



\begin{table}[H]
	\caption{Evaluation of the Malaria Parasite Classifier}
	\centering
	\begin{tabular}{lll}
		\midrule
		\bf{Model}     & \bf Accuracy     & \bf \#params \\
		\midrule
		$ResNet-50$ \cite{bib32} & 95.9\% & {25.6M}  \\
		$Xception$ \cite{bib32}& 89.0\%  & {22.8M} \\
		$AlexNet$ \cite{bib32}& 93.7\%  & {62.3M}  \\
		\textbf{MAISCOPE MobileNetV2 Model} & \textbf{89.9\%} &\textbf{2.2M}  \\
		\bottomrule
	\end{tabular}
	\label{tab:table}
\end{table}

{\subsection*{Recognition Pipeline Performance on Real-World Data}} 
Figure 10 shows MAIScopes model pipeline being tested with real world blood slides from India with blood cells that are infected with malaria. This image was provided by Cancyte Labs and the model was able to identify the two malaria infected cells with high confidence and identify where they are located. During training, the two models were never exposed to an image like this yet are still able to perform with high accuracy, highlighting how MAIScopes computer vison models can be applied to real world field data with high precision all while running on low-end hardware. 

\begin{figure}[h]
\centering
\includegraphics[width=0.8\textwidth]{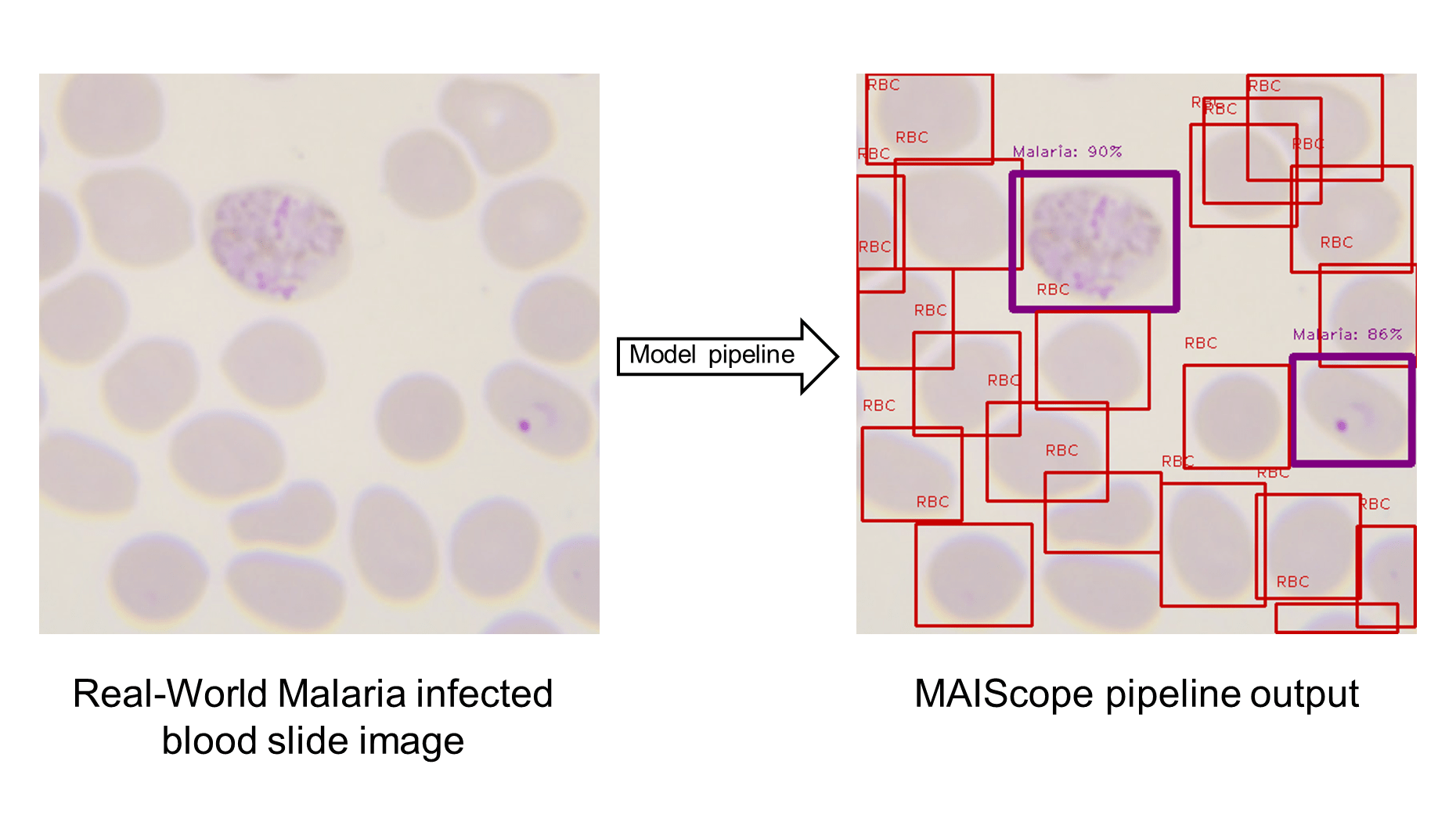}
\caption{Real malaria slide inputted into MAIScope model pipeline}
\label{fig:real_slide}
\end{figure}

\vspace{8pt}
\subsection*{{Cost of Materials }}
{Table 3 shows}
 the price breakdown of MAIScope. The minimum {total }price based on the retail price of components comes to about \$99.51 for a single MAIScope. Further optimization, sourcing, and bulk discounting can significantly reduce this price. This price meets the goals set at the beginning of the project which was to design a device that can be manufactured for less than \$125. Other methods of automated malaria and Disease diagnosis use mobile phones attached to the eye piece of a bulky light microscope which can cost upwards of \$1000. Alternatives are not only more expensive, but also are also not practical or portable. Assuming a cost of \$1100 for a screening configuration of a conventional light microscope paired with a mobile phone, 10 MAIScopes could be manufactured meaning 10x more malaria diagnosis output for the same price, an almost one-tenth reduction in cost. Not only is MAIScope cheaper, {but also} it solves the practical challenges of malaria diagnosis in rural areas which are not possible with alternative methods while still producing consistent results.

\begin{table}
	\caption{MAIScope Price Breakdown}
	\centering
	\begin{tabular}{ll}
		\midrule
		\bf Name of component    & \bf Retail price \\
		\midrule
		Raspberry Pi 4 Model B(2GB) & \$35.00 \\
		5” LCD display & \$20.00-\$30.00  \\
		8MP Camera & \$18.00-\$25.00  \\  
		5V Battery Pack & \$12.00-\$25.00 \\ 
        16mm Camera lens & \$8.00 \\ 
        Screws and 3D printed parts & $\simeq$ \$4.00 \\ 
        LED  & $\simeq$ \$2.50 \\ 
        Bead Microscope & \$0.01 \\ 
        \bf{Total} & $\simeq$ \$99.51-\$124.51 \\ 
		\bottomrule
	\end{tabular}
	\label{tab:table}
\end{table}

\subsection*{Evaluation of MAIScope’s Portability}
The final design of MAIScope can fit in a box with dimensions of 20x10x8{cm}. It does not require connection to a standard outlet but rather a portable battery pack that outputs 5V of energy. MAIScope also does not require access to the {I}nternet
to diagnose a patient with malaria as the software implementation is all offline. MAIScope meets all the requirements for portability and can be taken to rural areas without internet or stable access to power. 

\subsection*{Testing Embedded Bead Microscope Quality}
{Fig. 11 }shows a human blood slide under a conventional light microscope on the left (AmScope B120C-E1 Light Microscope) and the embedded bead microscope used for MAIScope on the right. The bead microscope comes in at a fraction of the cost while still showing an image that is comparable to the Light microscope, with individual red blood cells being identifiable and the loss in sharpness between the two images being minimal.  This satisfies the requirement set at the beginning of the project which was to use a microscope mechanism that is clear while eliminating the need for an expensive, bulky light microscope.

\begin{figure}[!h]
\centering
\includegraphics[width=0.8\textwidth]{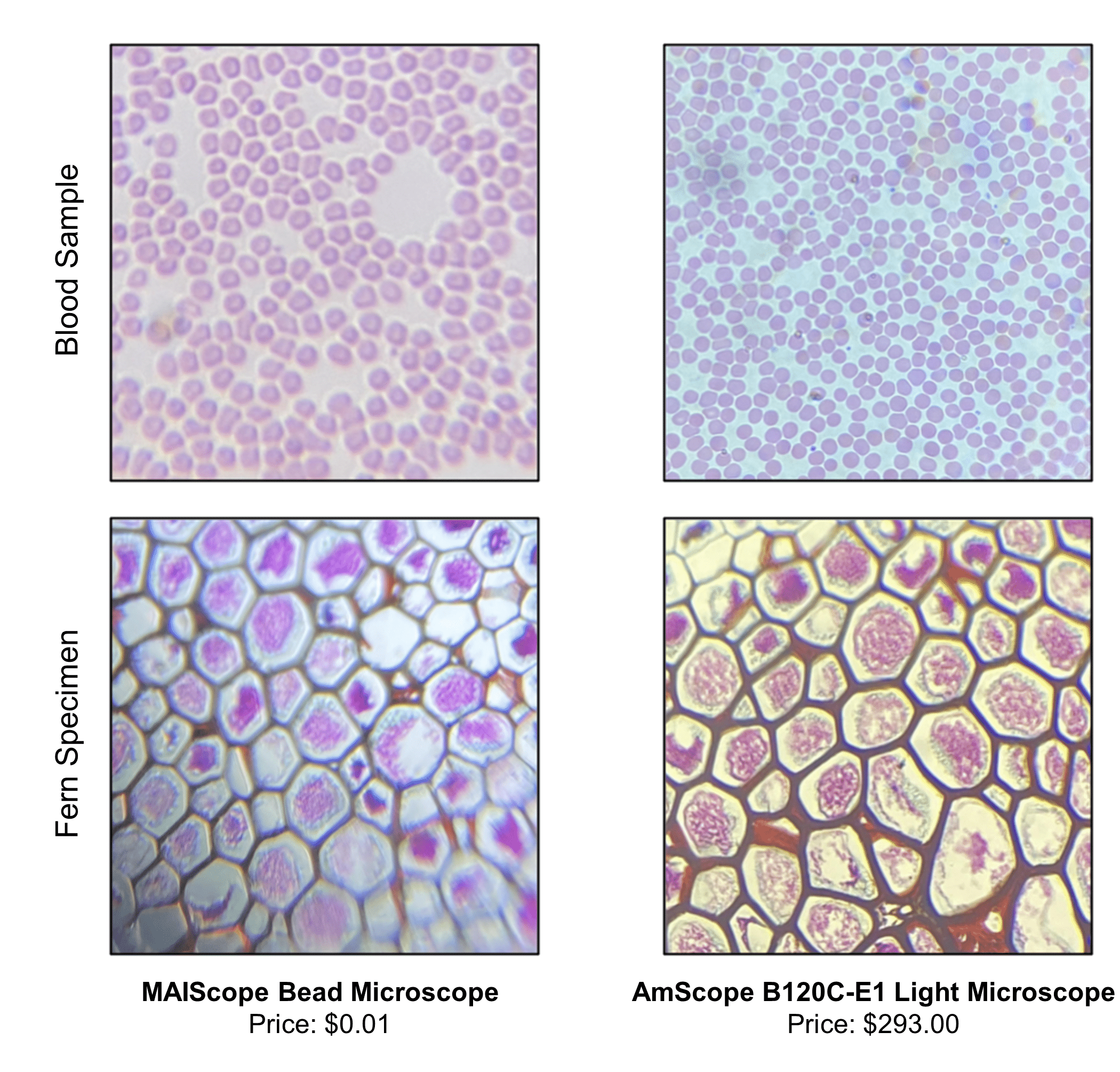}
\caption{Comparison between MAIScope bead microscope and conventional light microscope}
\label{fig:compare}
\end{figure}

\section{Discussion, Limitations and Next Steps}
MAIScope is a great step towards automating malaria microscopy in rural areas lacking resources. The portability and low-cost nature of the hardware coupled with its automatic {malaria parasite detection and recognition capabilities help} 
with timely diagnosis and treatment where expertise and equipment are {limited.} 

One big area the healthcare industry is lagging on is the availability of blood slide microscopic image data especially from the field and rural areas. This kind of data is critical because real world field data would include noise and variations present in the real world. For example, there will be variation in staining techniques which affect the color of the image. Similarly, contaminants such as dirt {may be present, which can alter the quality of the images}. 
{Furthermore, } variations in microscope equipment produce different image quality e.g., bead microscopes {as-compared-with} expensive light microscopes. One of the {future} goals of MAIScope is to save and collect all blood slides being scanned in the field. This will automatically be stored in a central cloud database and serve as an important asset for training the AI models of the future and help with the issue of data scarcity.

The next step for MAIScope is to go through more real-world field testing in India conducted by malaria experts in the lab. This will be an iterative process where the data collected as part of this testing will be used to retrain the machine learning models so it adapts to field variations. 
{A long term goal is to establish field trials with NGOs in India.}

{Since malaria often occurs in tropical locations, another future goal is to make the device more robust so it is usable in}
rough conditions {by} waterproofing for humid and wet environments.
Once the solution becomes robust, practical, and sturdy, it can be more widely adopted by national governments in nations that have a hard time treating malaria within their population.

\subsection*{Future Work}
MAIScope is designed such that it can go beyond malaria diagnosis. It can be adopted to other diseases typically diagnosed via microscopy. The approach is not limited to blood smears either. For example, scenarios where tissues are analyzed under the microscope (e.g.{, cancerous cells}) 
can also be automated with MAIScope. The field data collection goals 
{can}
also accordingly expand to support other disease diagnostics datasets.


\section{Conclusion}

MAIScope is a cohesive solution to tackle the problem of microscopic infecteous diseases in rural communities where laboratorians with expertise in malaria diagnosis are nonexistent and where healthcare access is limited. {The main objectives in designing MAIScope were \textbf{portability} and \textbf{simplicity} which makes diagnosis testing and treatments a lot more practical for medical practitioners in rural areas. To meet these objectives, MAIScope applies}
machine learning and computer vision approaches along 
{with low-end} hardware, lowering the overall cost of malaria diagnosis while still providing accurate results. 
MAIScope's {hardware design} consists of a Raspberry Pi 4, a Camera with 10x zoom, a touchscreen display and a Bead microscope system
{which}
does not require internet and only needs a battery pack with a 5V output. All of these components fit in a 20x10x8cm box and remove the need for bulky equipment, enabling MAIScope to be taken to the most rural 
areas.  The custom glass bead microscope utilized by MAIScope is very cheap but provides very similar imaging results to that of an expensive conventional light microscope which can cost hundreds of dollars. In order to provide accurate malaria diagnosis, MAIScope has two state of the art TensorFlow Lite{-based} models capable of producing precise results while still being able to run on a Raspberry Pi 4. The first model uses the BCCD dataset and is a blood cell detection model which can identify three different types of cells in blood: white blood cells, red blood cells, and platelets. This model outperformed 3 other models designed to run-on high-end hardware trained on the same dataset while being a fraction of the size. The second model MAIScope uses is a custom malaria classification model which was trained on a malaria dataset. This model is able to identify if a blood cell is infected with malaria with 89.9\% accuracy, which is just a little lower than other malaria classification models designed for high end expensive hardware. These two models were combined into a pipeline in which red blood cells extracted from an input image of a blood slide were sent as input to the malaria classification model. At the very end of the pipeline the number of infected cells are counted to provide insight on the severity of a malaria infection and the results are displayed to the user on the touch screen display. The amount of user interaction involved with MAIScope is limited to avoid human error. All the user has to do is insert a prepared blood slide and take a picture using MAIScope’s easy to use UI and repeat until done with the screening session.

Future work includes waterproofing MAIScope 
{as well as}
adapting the device for other disease diagnosis that require microscopy. Other work includes field testing MAIScope in rural settings, and further lowering costs. One limitation in this work was the fact that we did not have direct access to  malaria blood slides, only images so it is a future goal to test MAIScope with physical malaria slides and see how it performs.  


\section{Acknowledgements}
I am very thankful to Dr. Sridhar, a practicing surgeon and urologist at Rangadore Hospital, Bengaluru, India and Managing Director and Chairman at Cancyte, a research institute dedicated to researching and providing cost effective and affordable medical solutions in India. He is well known in India for playing an active role in stopping the spread of infectious diseases like Dengue, Chikungunya and others in rural parts of India. Dr. Sridhar, being a very busy individual, was very generous with his time to educate me on malaria parasites and the real-world challenges in rural developing nations. Dr Sridhar’s mentorship has significantly bolstered my passion for AI driven disease diagnosis, which culminated from my own health issues that took a long time to diagnose. I am also thankful to the scientists at Cancyte Labs for answering my questions regarding malaria parasite and blood smearing techniques used in rural areas.

\end{document}